\documentclass[a4paper,12pt]{article}

\usepackage{amssymb,latexsym}

\makeatletter \@addtoreset{equation}{section} \makeatother

\begin{document}

\begin{titlepage}

\thispagestyle{empty}

\begin{flushright}
\hfill{hep-th/0511122}
\end{flushright}

\vspace{35pt}

\begin{center}{ \Large{\bf
Supersymmetric compactifications of heterotic strings with
fluxes and condensates }}

\vspace{60pt}

{\bf  Pantelis Manousselis}$^1$,\ {\bf Nikolaos Prezas}$^2$ {\bf and} {\bf George Zoupanos}$^3$  \\
\vspace{25pt}
$^1${\it Department of Engineering Sciences, University of Patras,\\
GR-26110 Patras, Greece\\

{\tt pantelis@upatras.gr}} \vspace{10pt}


$^2${\it Institut de Physique, Universit\'e de Neuch\^atel, \\
CH--2000 Neuch\^atel, Switzerland \\

{\tt nikolaos.prezas@unine.ch}}

\vspace{10pt}

$^3${\it Physics Department, National Technical University of Athens, \\
GR-15780 University Campus, Athens, Greece \\

{\tt zoupanos@mail.cern.ch}}

\vspace{20pt}

{ABSTRACT}

\end{center}

\vspace{20pt}

We discuss supersymmetric compactifications of heterotic strings in
the presence of  H-flux and general condensates using the formalism
of $G$-structures and intrinsic torsion. We revisit the examples
based on nearly-K\"ahler coset spaces and show that supersymmetric
solutions, where the Bianchi identity is satisfied, can be obtained
when both gaugino and dilatino condensates are present.


\vspace{20pt}

\end{titlepage}

\newpage

\baselineskip 6 mm
\section{Introduction}

In string theory compactifications on Calabi-Yau  (CY) manifolds and
the corresponding dimensional reductions \cite{Candelas:1985en}, the
resulting low-energy field theory in four dimensions typically
contains a number of massless chiral fields, characteristic of the
internal geometry, known as moduli. These fields, arising in four
dimensions as massless modes of the higher-dimensional matter fields
and from the gauge-independent variations of the metric of the
compact space, correspond generically to flat directions of the
four-dimensional effective potential. Therefore, the values taken by
the moduli in the vacuum, which in turn specify the masses and
couplings of the four-dimensional theory, are left undetermined.
Hence, the theory is without predictive power.

Fortunately, the moduli problem in the form described above appears
only in the simplest choice of string backgrounds, where out of the
plethora of  closed-string fields only the metric is assumed to be
non-trivial. By considering more general backgrounds involving
``fluxes" \cite{Strominger:1986uh, deWit:1986xg} as well as
non-perturbative effects \cite{Dine:1985rz, Derendinger:1985kk}, the
four-dimensional theory can be provided with potentials for some or
all moduli. The terminology ``fluxes" refers to the inclusion of
non-vanishing field strengths for the ten-dimensional antisymmetric
tensor fields with directions purely inside the internal manifold.
Therefore, the present day problem is the choice of the appropriate
background which could lead to realistic and testable
four-dimensional theories.

The presence of fluxes has a dramatic impact on the geometry of the
compactification space. Specifically, the energy carried by the
fluxes back-reacts on the geometry of the internal space and the
latter is deformed away from Ricci-flatness. Then, the CY manifolds
used so often in string theory compactifications cease to be  true
solutions of the theory. For example, the requirement that some
supersymmetry is preserved implies that the internal manifold is a
non-K\"ahler space for heterotic strings with NS-NS fluxes
\cite{Cardoso:2002hd,Curio:2000dw, Becker:2003yv,Becker:2003sh}, while it can be
a non-complex manifold for type IIA strings
\cite{Dall'Agata:2003ir,Behrndt:2004mj, Lust:2004ig}. For type IIB
strings, instead, the deviation is mild since the overall effect due
to the fluxes is a conformal rescaling of the original CY solution
\cite{Giddings:2001yu, Dall'Agata:2004dk}.
 We should mention at this point that in principle we can consider
compactifications on
CY manifolds with fluxes too, but these are reliable only in the
large-volume limit where the flux back-reaction can be
consistently ignored.

A considerable amount of literature has been devoted to the problem
of including appropriately the back-reaction of the fluxes on the
internal manifold and constructing examples of manifolds which are
true solutions of the theory. In general, these manifolds have
non-vanishing torsion. Consequently, demanding that the
low-energy theory is supersymmetric implies that the internal
manifold admits a $G$-structure \cite{Gauntlett:2003cy}. The existence
of a $G$-structure is
a generalization of the condition of special holonomy. Subsequently,
the allowed  $G$-structures can be classified in terms of their
intrinsic torsion classes \cite{grey}.

For heterotic strings, requiring that supersymmetry is preserved in
the presence of non-perturbative effects such as gaugino
condensation \cite{Cardoso:2003sp}, leads to  $AdS_{4}$
spacetimes with non-complex internal
manifolds as potential solutions \cite{Frey:2005zz} \footnote{Originally,
 gaugino condensation had been suggested
as a supersymmetry breaking mechanism in a Calabi-Yau
compactification \cite{Dine:1985rz,Derendinger:1985kk}. Here,
instead, we consider supersymmetric solutions on manifolds that are
not Calabi-Yau.}. The non-complex manifolds that we will consider
here are simple homogeneous nearly-K\"ahler coset spaces. They were
identified as interesting possible solutions of heterotic string
theory in refs.~\cite{Castellani:1986rg, Lust:1986ix}, whereas
supersymmetric solutions of the form $AdS_{4} \times G/H$ were first
obtained in refs.~\cite{Govindarajan:1986iz, Govindarajan:1986kb}.
In order to obtain such solutions the authors in
refs.~\cite{Govindarajan:1986iz, Govindarajan:1986kb} assumed the
presence of a gaugino condensate and  performed a case by case
analysis. More recently, the 10-dimensional supersymmetry conditions
in the presence of a non-vanishing gaugino condensate were examined
in the language of $G$-structures in ref.~\cite{Frey:2005zz}.

Here, we generalize the setup of ref.~\cite{Frey:2005zz} by
considering more exotic condensates. The consideration of other
condensates besides that due to the gaugino is imposed upon us for
both technical and aesthetical reasons. At a technical level, the
source-free Bianchi identity cannot be solved in a
supersymmetry-preserving manner if only the gaugino condensate is
present. One of the objectives of the present work is to show that
supersymmetric solutions of the Bianchi identity can be obtained if,
in addition to the gaugino condensate, a dilatino condensate acquires
a non-vanishing vacuum expectation value (vev).
On aesthetic grounds and although there is no known
mechanism leading to condensation of the dilatino \footnote{See
however refs.~\cite{Konishi:1988mb,Konishi:1989em} where
condensation of fermions in the gravity sector is considered in a
different context and ref.~\cite{Kitazawa:2001zi} which discusses a
similar effect in a 5-brane background. Moreover,
ref.~\cite{Duff:1982yi} considered Minkowski vacua in 11-dimensional
supergravity with gravitino condensates.}, we adopt here a more broad
point of view which considers the condensates on equal footing with
the fluxes. Following this approach,  we first formulate the
supersymmetry conditions in the formalism of $G$-structures and
then we revisit the solutions
of refs.~\cite{Govindarajan:1986iz, Govindarajan:1986kb}, while
completing also their list by one more example. 

We should mention
that, as in most works that study 
the supersymmetry constraints on the geometry,
we do not check explicitly
the equations of motion. In our case this would be a non-trivial
task because the presence of the condensates renders the derivation
of the equations of motion subtle \cite{Frey:2005zz}.
Instead, we rely on the fact that  backgrounds that preserve
some supersymmetry and where the Bianchi equations and the equations of motion for
the matter fields are satisfied, are automatically 
solutions of the Einstein
equations \cite{Lust:2004ig, Gauntlett:2002fz}. 
It is straightforward to show that this is indeed true for
hererotic strings by specializing the
results of  \cite{Gauntlett:2002fz} to hererotic M-theory.

A key feature  of the solutions we consider is that since they are
not Ricci-flat, the four-dimensional spacetime will not be Minkowski
but $AdS$ (at least in the case where some supersymmetry is preserved). 
For this reason, the Ricci-flat CY manifolds were
originally more attractive candidates. Recently, however, this has
become less relevant since whenever the inclusion of fluxes produces
a stable vacuum without moduli, this vacuum turns out to be anti-de Sitter.
Therefore, it is not a serious drawback to start from an
anti-de-Sitter solution in the first place and hope that eventually
non-perturbative effects will lift this to a Minkowski or de Sitter
vacuum according, for example, to the scenario proposed in
\cite{Kachru:2003aw} (see also \cite{Curio:2001qi} 
for an explicit construction of this type in heterotic M-theory).
This process may result in interesting GUTs in
four-dimensions \cite{Burgess:2003ic,Lust:1985be,
Manousselis:2004xd,Manousselis:2001re,Manousselis:2001xb,Manousselis:2000aj}.

The layout of this paper is as follows. In section 2 we establish
our notation and we present the supersymmetry transformations of the
fermionic fields of heterotic supergravity when various condensates
are non-vanishing. In section 3 we examine the conditions imposed on the
external and internal geometries under the requirement of preserving
${\cal N}=1$ supersymmetry in the external space. The conditions on
the internal geometry are formulated in terms of the intrinsic
torsion classes of the $SU(3)$-structure. In section 4 we study some
specific solutions based on nearly-K\"ahler coset spaces. We show
that the Bianchi identity can be satisfied if and only if both
gaugino and dilatino condensates are present. Finally, in section 5
we present a few concluding remarks. Also, in two appendices we provide our
Gamma matrix conventions and some details on the $SU(3)$-structure
of the cosets under consideration.

\section{Heterotic strings with condensates}

The fields of heterotic supergravity, which is the low-energy limit
of heterotic superstring theory, consists of the  ${\cal N} =1,
D=10$ supergravity multiplet which contains the fields $e_{M}^N,
\psi_{M}, B_{MN}, \lambda, \varphi$, (i.e.~namely the metric, the
gravitino which is a Rarita-Schwinger field, the two-form
potential, the dilatino which is a Majorana-Weyl spinor, and the
dilaton which is a scalar), coupled to a ${\cal N} =1,  D=10$ vector
supermultiplet which contains
the gauge field $A_{M}$ and the corresponding gaugino
$\chi$. The gauge field and the gaugino transform in the
adjoint of $E_{8} \times E_{8}$. The Lagrangian and the full
supersymmetry variations can be found in
refs.~\cite{Bergshoeff:1981um,Chapline:1982ww,Bergshoeff:1989de}.

The supersymmetry variations of the fermionic fields, including the
relevant for our discussion fermion bilinears, are
\cite{Chapline:1982ww}\footnote{Note that the coefficient of the
last term in $\delta\lambda$ was corrected in
ref.~\cite{Dine:1985rz}.}
\begin{eqnarray}
\delta\psi_{M} &=& \nabla_{M}\epsilon +
\frac{\sqrt{2}}{32}\varphi^{-3/4}
\left(\Gamma_{M}\Gamma^{NPQ}-12\delta_{M}^{N}\Gamma^{PQ}\right)\hat H_{NPQ}\epsilon
\nonumber \\ &-& \frac{1}{256}\left(\Gamma_{M}\Gamma^{NPQ} -
8\delta_{M}^{N}\Gamma^{PQ}\right) \Big({\rm Tr}(\bar\chi\Gamma_{NPQ}\chi)
+\frac{1}{2}(\bar\lambda \Gamma_{NPQ}\lambda)
\Big)
\epsilon +\frac{\sqrt{2}}{96}
(\bar\psi_M \Gamma_{KL}\lambda)\Gamma^{KL}
\epsilon,\nonumber\\
\delta\chi &=& -
\frac{1}{4}\varphi^{-3/8}\Gamma^{MN}\hat F_{MN}\epsilon+\frac{\sqrt{2}}{64}
\Big(3 (\bar \lambda\chi)
-\frac{3}{2}(\bar \lambda\Gamma_{MN}\chi)
\Gamma^{MN}
-\frac{1}{24} (\bar \lambda\Gamma_{MNKL}\chi)
\Gamma^{MNKL}\Big)\epsilon
, \nonumber\\
\delta \lambda &=& -
\frac{3\sqrt{2}}{8}\varphi^{-1}\Gamma^{M}\partial_{M}\varphi\epsilon
+ \frac{1}{8}\varphi^{-3/4}\Gamma^{MNP}\hat H_{MNP}\epsilon +
\frac{\sqrt{2}}{384}{\rm Tr}(\bar\chi\Gamma_{MNP}\chi)
\Gamma^{MNP}\epsilon \label{origsusy}.
\end{eqnarray}
We have set the gravitational and Yang-Mills coupling constants
equal to 1. This in turn implies that we work with units where
$\alpha'=4$. The supersymmetry parameter $\epsilon$ is a
10-dimensional Majorana-Weyl spinor with 16 real components. The
hats denote the supercovariant generalization of the corresponding
fields,
\begin{eqnarray}
\hat F_{MN}&=&F_{MN} - \varphi^{3/8} (\bar \psi_{[M} \Gamma_{N]} \chi), \\
\hat H_{MNK}&=&H_{MNK}-\frac{1}{4}  \varphi^{3/4}\Big( \sqrt{2}
(\bar\psi_{[M} \Gamma_{N}\psi_{K]}) - (\bar \psi_{[M} \Gamma_{NK]}
\lambda)\Big).\label{sch}
\end{eqnarray}

It is well-known that at the supergravity level the gauge-invariant
3-form field strength is $H_{MNP}=3 \partial_{[M}
B_{NP]}-\frac{1}{\sqrt{2}}A_{[M}F_{NP]}$. The corresponding Bianchi
identity reads as
\begin{equation}\label{eq:2.4}
dH=- \frac{\sqrt{2}}{6} {\rm tr} (F\wedge F).
\end{equation}
The full Bianchi identity includes one more term that is a
string-theoretic correction and will be added later.

In the above supersymmetry transformations we have assumed that some
of the possible condensates between the fermionic fields of
heterotic string theory have non-vanishing vacuum expectation
values. Our motivation for keeping only those appearing above is
that, as we will see later, they permit supersymmetric solutions
without rendering the analysis computationally challenging. We will
also assume that the condensates $(\bar \psi_{[M} \Gamma_{N]} \chi)$
and $(\bar\psi_{[M} \Gamma_{N}\psi_{K]})$, which appear in the
supercovariant field strengths, are  vanishing. We postpone the
presentation of a more complete analysis incorporating all possible
condensates for future work \cite{wip}.

An issue  that should be addressed at this point concerns the supersymmetry
variations of the bosonic fields in the presence of fermion condensates.
These read \cite{Chapline:1982ww}
\begin{eqnarray}
\delta e_M^L&=&\frac{1}{2} \bar\epsilon  \Gamma^L \psi_M,\;\;\;\;\;
\delta\varphi=-\frac{1}{3}\sqrt{2} \epsilon\lambda \varphi,\;\;\;\;\;
\delta A_M=\frac{1}{2}\varphi^{3/8}\bar\epsilon\Gamma_M\chi,\nonumber\\
\delta B_{MN}&=&\frac{1}{4}\sqrt{2} \varphi^{3/4} (\bar\epsilon
\Gamma_M\psi_N-\bar\epsilon \Gamma_N \psi_M-\frac{1}{2}\sqrt{2}\epsilon
\Gamma_{MN}\lambda)+\frac{1}{2}\sqrt{2}\varphi^{3/8}\epsilon\Gamma_{[M}{\rm Tr}
(\chi A_{N]})
\end{eqnarray}
and it is obvious that had the fermions acquired non-zero vevs,
these variations would not vanish in general. In such a case, of course,
checking the supersymmetry variations would be redundant 
since the non-vanishing vevs for the fermions would imply anyway
broken spacetime (e.g.~Lorentz) symmetry.
The crucial point here is that in quantum field theory
the fermion condensates can be non-zero while maintaining
a vanishing fermion vev. Therefore, although the fermion condensates
are assumed to be non-vanishing, the vevs of the corresponding fermions are taken to
be zero, as is necessary for preserving maximal symmetry 
in spacetime. It should be emphasized that this effect has been
explicitly demonstrated 
in ${\cal N}=1$ Super-Yang-Mills theory where
gaugino condensation indeed takes place non-perturbatively 
without leading to non-zero
vevs for the gauginos \cite{Novikov:1983ee, Novikov:1983ek}.
Here, we assume that this is precisely what happens
in the strongly-coupled 
microscopic theory that underlies our effective supergravity
description.
This is in line with the original treatment of gaugino condensation
in hererotic supergravity where the fermion vevs are zero
while a non-trivial condensate is generated
\cite{Dine:1985rz, Derendinger:1985kk}. 
The punchline is that the supersymmetry variations of the bosons
are trivially vanishing since the fermion vevs are consistently zero.

In our setup, the only extra assumption
is that some unknown quantum effects can  lead  to non-trivial
condensates for the gravitinos. Actually, such a point of view was
already taken in \cite{Duff:1982yi} while \cite{Kitazawa:2001zi} provided
some evidence that this effect indeed takes place in NS5-brane backgrounds.
Furthermore, we should stress that here we simply work in an effective approach where we ask
what would happen {\em if} such condensates were generated.
Hence, our treatment is entirely analogous to the usual
chiral Lagrangian approach to hadron physics. In this approach,
one simply assumes that in the IR a non-vanishing quark condensate breaking
chiral symmetry is formed, although the miscoscopic theory (i.e.~QCD) 
governing this effect is out of reach in this regime. Moreover, in order
to maintain Lorentz invariance
the quark vevs are assumed to be zero despite the presence
of the non-perturbative quark condensate.

We now make the following field redefinitions:
\begin{eqnarray}
\varphi &=& e^{-8/3\phi}, \ \ \ \ \ \ \ \ \ \ g_{MN} = e^{-2\phi}
g_{MN}^{(0)}, \nonumber \\
\lambda &=& \frac{1}{\sqrt{2}} e^{\phi/2}\lambda^{(0)}, \ \ \ \ \
\psi_{M} = e^{-\phi/2}(\psi^{(0)}_{M} -
\frac{\sqrt{2}}{4}\Gamma^{(0)}_{M}\lambda^{(0)}), \nonumber \\
\epsilon &=& e^{-\phi/2}\epsilon^{(0)}, \ \ \  \ H_{MNP} =
\frac{3}{\sqrt{2}}H^{(0)}_{MNP}, \nonumber \\
\chi &=& e^{\phi/2} \chi^{(0)}, \ \ \ \ \ \ \ F_{MN} = F_{MN}^{(0)}
\nonumber, \\ \Gamma_{M} &=& e^{-\phi}\Gamma^{(0)}_{M},
\end{eqnarray}
with the quantities bearing the superscript $(0)$
referring to those in (\ref{origsusy}). The supersymmetry variations
(\ref{origsusy}) become
\begin{eqnarray}
\delta\psi_M&=&\nabla_M
\epsilon-\frac{1}{4}\Big(\hat H_{M}-2\Sigma_{M}-\frac{4}{3}\Delta_M\Big)\epsilon
-\frac{1}{4}\Gamma_M\Big(\frac{1}{3}
\Sigma+\frac{1}{4}\Delta\Big)
\epsilon, \label{susy1}\\
\delta\chi&=&-\frac{1}{4}\Gamma^{MN} \hat F_{MN}\epsilon +\frac{1}{32}
\Big(3 \Phi-\frac{3}{2}\Phi_{MN}\Gamma^{MN}-\frac{1}{24}\Phi_{MNKL}
\Gamma^{MNKL}\Big)\epsilon,
\label{susy2}\\
\delta\lambda&=&\nabla\!\!\!\!\slash\phi\epsilon+\frac{1}{24}\Big(\hat H
+ \Sigma\Big)\epsilon,\label{susy3}
\end{eqnarray}
where $\Phi_{[\ldots]}=(\bar \lambda \Gamma_{[\ldots]}\chi)$. Notice that
since these condensates are not singlets of the $E_{8} \times E_{8}$ gauge group,
turning on vevs for them can break part of the original gauge symmetry. Of course,
this is similar to the (partial) breaking of the original
gauge symmetry due to a non-vanishing background gauge field strength.

In the field-redefinitions above we set $(\bar \lambda \Gamma_M
\Gamma_{KL}\lambda)=(\bar \lambda\Gamma_{MKL}\lambda)$ because the
difference involves terms of the form $(\bar \lambda \Gamma_M
\lambda)$ which are zero due to the assumption of maximal symmetry
in the external space and due to the absence of globally-defined
vector fields on six-dimensional manifolds with $SU(3)$-structure.
In the new variables the Bianchi identity (\ref{eq:2.4}) reads as
\begin{equation}
dH=- \frac{1}{2} {\rm tr} (F\wedge F),
\end{equation}
with the  supercovariant field strengths (\ref{sch})  being
\begin{eqnarray}
\hat F_{MN}&=&F_{MN}, \\
\hat H_{MNK}&=&H_{MNK}-6 \Delta_{MNP}, \label{schnew}
\end{eqnarray}
since after the field redefinitions we set  $(\bar
\psi_{[M} \Gamma_{NK]} \lambda)=0$.
We use the standard shorthand notation $H_{M} = H_{MNP}\Gamma^{NP}$,
$H= H_{MNP}\Gamma^{MNP}$, $\Sigma_{M} = \Sigma_{MNP}\Gamma^{NP}$,
$\Sigma = \Sigma_{MNP}\Gamma^{MNP}$ with $\Sigma_{MNP}=\frac{1}{16}
{\rm tr}(\bar \chi \Gamma_{MNP}\chi)$ and
$\Delta_{MNP}=\frac{1}{16}(\bar \lambda \Gamma_{MNP}\lambda)$.

We emphasize that the metric in the above equations is the {\em sigma-model} metric and
is related to the Einstein metric $g^0_{MN}$ as $g_{MN}=e^{-2\phi}
g^0_{MN}$. We have also used the fact that
the covariant derivatives of a spinor with respect to
these two metrics are related as
$\nabla_{M}\epsilon = \nabla_{M}^{(0)}\epsilon -
\frac{1}{2}\Gamma_{M}^{\;\;\;N}\partial_{N}\phi\epsilon $.

\section{Conditions for 4-dimensional ${\cal N}=1$ vacua}

\subsection{Metric ansatz and $SU(3)$-structure}

We assume that $g^0_{MN}$ describes the warped product
$M_{1,3}\times_w K_6$ of a 4-dimensional maximally symmetric
spacetime $M_{1,3}$ with
a compact 6-dimensional internal space $K_6$.
Explicitly
\begin{equation}
ds^2=g^0_{MN}(x,y) dx^M dx^N=e^{2D(y)} \Big(\hat{g}_{\mu\nu}(x)
dx^\mu dx^\nu+ \hat{g}_{mn}(y) dy^m dy^n\Big).
\end{equation}


Since we are interested in 4-dimensional vacua with
${\cal N}=1$ supersymmetry, we demand that on $K_6$ there exists
a globally-defined complex spinor $\eta_+$
(and its conjugate $\eta_{-}$ with opposite chirality). In other words
 $K_6$ should be  equipped with an $SU(3)$-structure.
Our ansatz for the 10-dimensional Majorana-Weyl spinor $\epsilon$ is
\begin{equation}
\label{spinans} \epsilon(x,y)=f(y) \theta_+(x)\otimes\eta_+(y) -
f^*(y)\theta_-(x)\otimes\eta_-(y),
\end{equation}
where $f(y)$ is an arbitrary complex function. This ansatz yields
${\cal N}=1$ supersymmetry in $M_{1,3}$ expressed in terms of the
Weyl spinors $\theta_\pm$. The 6-dimensional spinors are
normalized\footnote{Our spinor conventions and some useful formulae
can be found in Appendix A.} as
$\eta^\dagger_+\eta_+=\eta^\dagger_-\eta_-=1$. As usual the
four-dimensional spinors are taken to be Grassmann while the
six-dimensional ones are commuting.

The $SU(3)$-structure is characterized by an almost complex
structure and\footnote{As is common, we will use the same symbol for
both the almost complex structure tensor and the associated 2-form.}
the associated 2-form $J_{mn}$, and by a (3,0)-form $\Omega_{mnp}$.
These forms are globally-defined and non-vanishing and they are
subject to the following compatibility conditions:
\begin{equation}
J \wedge \Omega=0, \;\;\; \Omega \wedge \Omega^* = \frac{4 i}{3} J\wedge J\wedge J.
\end{equation}


Furthermore, the spinors $\eta_\pm$ determine (up
to a phase which we fix to a convenient value) the $SU(3)$-structure
forms as
\begin{eqnarray}
J_{mn}&=&-i \eta^\dagger_+ \gamma_{mn} \eta_+=i \eta^\dagger_- \gamma_{mn} \eta_-,\\
\Omega_{mnp}&=&\eta^\dagger_-\gamma_{mnp}\eta_+,\label{omega} \\
\Omega^*_{mnp}&=&-\eta^\dagger_+\gamma_{mnp}\eta_- .
\end{eqnarray}
Using Fierz identities one can show that indeed $J_m^{\;\;n}$
is an almost complex structure, i.e.~it satisfies
$J_m^{\;\;n} J_n^{\;\;p}=-\delta_m^p$. Then,  the projectors
$(\Pi^\pm)_m^{\;\;n}=\frac{1}{2}(\delta_m^{\;\;n}\mp i
J_m^{\;\;n})$ can be used to separate the holomorphic
and antiholomorphic parts of a generic form.

Now we can  decompose the condensates in 4- and 6-dimensional
pieces. First, notice the gaugino field $\chi$ is Majorana and of
the same chirality as the supersymmetry parameter $\epsilon$. Hence,
it admits a decomposition $\chi=\psi_+\otimes\eta_+
-\psi_-\otimes\eta_-$ with $\psi_{\pm}$ the 4-dimensional gaugino
fields. Notice that this is not a zero-mode decomposition but
the usual decomposition in terms of $SU(3)$ singlets. 
Then we obtain
\begin{equation}
\Sigma_{mnp}= \frac{1}{16}{\rm Tr}(\bar\chi
\Gamma_{mnp}\chi)=-\Big(\psi^\dagger_-\psi_+ \Omega_{mnp}
 +\psi^\dagger_+\psi_-\Omega^*_{mnp}\Big)=-(\Lambda^3 \Omega_{mnp}+
 c.c.),
\label{cond}
\end{equation}
where $\Lambda^3=\frac{1}{2}{\rm
Tr}(\bar\psi(1+\gamma_{(5)})\psi)=\psi^\dagger_-\psi_+$ the
4-dimensional condensate. We see that the condensate consists only
of $(3,0)$ and $(0,3)$ pieces. This expansion is valid for
$\Delta_{mnp}$ as well. By denoting  $\delta^3=\frac{1}{2}{\rm
Tr}(\bar\lambda(1+\gamma_{(5)})\lambda$ the vev of the
4-dimensional dilatino condensate, we have $\Delta_{mnp}=-\delta^3
\Omega_{mnp}+ c.c.$

In order to expand correctly the dilatino-gaugino condensates
$\Phi_{[\cdots]}$ we have to take into account that $\chi$ and
$\lambda$ have opposite chiralities. Hence, $\lambda$ is expanded as
$\lambda=\lambda_+\otimes\eta_- -\lambda_-\otimes\eta_+$ with
$\lambda_{\pm}$ the 4-dimensional dilatinos. Then we find that
$\Phi$ is a real scalar $\Phi=-(\lambda_+^\dagger\chi_- +
\lambda_-^\dagger\chi_+)$, $\Phi_{mn}$ is a real 2-form $\Phi_{mn}=i
(\lambda_+^\dagger\chi_- - \lambda_-^\dagger\chi_+) J_{mn}\equiv
\Phi_0 J_{mn}$, and $\Phi_{mnkl}$ is a real 4-form $\Phi_{mnkl}=-3
\Phi J_{[mn} J_{kl]}$. Also, all of them have  an adjoint $E_8
\times E_8$ index that we suppress.

The supercovariant H-flux (\ref{sch}) can be expanded in terms of
the $SU(3)$-invariant forms as
\begin{equation}
\label{hflux} \hat H_{mnp}= \frac{1}{48} \Omega_{mnp} \hat
H^{(3,0)}+\Big(\hat H^{o(2,1)}_{mnp}+\frac{3}{4} \hat
H^{(1,0)}_{[m}J_{np]}\Big)+ c.c.,
\end{equation}
where $\hat H^{(3,0)}=\Omega^{*mnp}\hat H_{mnp}, \; \hat
H^{(1,0)}_m=(\Pi^+)_m^{\;\;s}\hat H_{snp} J^{np}$ and $\hat
H^{o(2,1)}$ the primitive $(2,1)$ piece of $\hat H_{mnp}$ which
satisfies $\hat H^{o(2,1)} \wedge J =0$. Notice that due to
(\ref{schnew}) and the fact that $\Delta_{mnp}$ consists of only
$(3,0)$ and $(0,3)$ pieces, we have $\hat
H_{mnp}^{(2,1)}=H_{mnp}^{(2,1)}$.

Now, we are ready to proceed to the analysis of the supersymmetry variations
(\ref{susy1}), (\ref{susy2}) and (\ref{susy3}) for the setup under
consideration.

\subsection{Conditions on the external geometry}

The 4-dimensional part of the gravitino variation reads as
\begin{equation}
\delta\psi_\mu=\nabla_\mu\epsilon-\frac{1}{12}
\tilde\Sigma_{mnp}\Gamma_\mu \Gamma^{mnp}\epsilon=0,
\label{4dgrav}
\end{equation}
where we have introduced the combination $\tilde
\Sigma_{mnp}=\Sigma_{mnp}+ \frac{3}{4}\Delta_{mnp}=-(\tilde\Lambda^3
\Omega_{mnp}+c.c.)$. Since $g_{MN}=e^{2(D-\phi)}\hat{g}_{MN}$ and
$\Gamma_M=e^{(D-\phi)}\hat{\Gamma}_M$ we can rewrite
eq.~(\ref{4dgrav}) as
\begin{equation}
\delta\psi_\mu=\hat{\nabla}_\mu\epsilon+\frac{1}{2}\hat{\Gamma}_\mu^{\;\;n}
\hat{\nabla}_n(D-\phi) \epsilon-\frac{1}{12}
e^{(D-\phi)}\hat{\Gamma}_\mu\tilde\Sigma\epsilon,
\end{equation}
and its integrability condition gives $\hat{\Gamma}^{\mu\nu}
\hat{\nabla}_\nu \delta\psi_\mu=0$ \cite{Strominger:1986uh}.

By assumption the 4-dimensional metric $\hat{g}_{\mu\nu}(x)$ is
maximally symmetric. Hence, the corresponding Riemann curvature tensor is
of the form
$$\hat{R}_{\nu\mu\kappa\lambda}=\frac{\hat{R}}{12}(\hat{g}_{\nu\kappa}\hat{g}_{\mu\lambda}-
\hat{g}_{\nu\lambda}\hat{g}_{\mu\kappa})$$ with $\hat{R}$ the {\em
constant} scalar curvature. For this metric we can show easily that
$\hat{\Gamma}^{\mu\nu}\hat{\nabla}_\nu\hat{\nabla}_\mu\epsilon=\frac{\hat{R}}{4}\epsilon$.
Then the integrability condition yields
\begin{equation}
\label{integra} \frac{\hat{R}}{4}-3
\hat{\nabla}^n(D-\phi)\hat{\nabla}_n(D-\phi)+ 3 \tilde\Sigma_m
\hat{\nabla}_m e^{(D-\phi)}-\frac{1}{12}e^{2(D-\phi)} \tilde\Sigma^2 =0.
\end{equation}
For vanishing condensates  eq.~(\ref{integra}) reduces to
eq.~$(2.8)$ of
\cite{Strominger:1986uh}. 
In this case one finds that $\hat{R}=0$
since the only possible constant value for
$\hat{\nabla}^n(D-\phi)\hat{\nabla}_n(D-\phi)$ on a compact manifold
is zero. Hence $M_{1,3}$ is Minkowski and the warp factor satisfies
$D(y)=\phi(y)$ up to an additive constant.

At first sight a non-zero
condensate $\tilde\Sigma_{mnp}$ seems to open up a much wider range of possibilities.
However, we will see in a moment that the supersymmetry conditions
imply the constancy of $D-\phi$. Then $M_{1,3}$ is $AdS_4$ with
constant negative curvature given by
\begin{equation}
\hat{R}=\frac{1}{3}e^{2(D-\phi)}\tilde\Sigma^2.
\label{4dcurv}
\end{equation}
Notice that $\tilde\Sigma^2$ is negative because $\tilde\Sigma$ is antihermitian.

Another way of computing the dependance of $\hat{R}$ on the
vev of the condensate $\tilde\Lambda$ is the following.
The 4-dimensional spinors in a maximally symmetric spacetime satisfy
$\hat\nabla_\mu \theta_+=W\hat\gamma_\mu\theta_-$ and
$\hat\nabla_\mu \theta_-=W^*\hat\gamma_\mu\theta_+$. The curvature
is then
$\hat{R}=-48 |W|^2$.
Using the fact that
$$\nabla_\mu\theta_+=\hat\nabla_\mu\theta_++\frac{1}{2}\Gamma_{\mu}\Gamma^m
\partial_m(D-\phi) \theta_+=We^{-(D-\phi)}\gamma_\mu\theta_-+
\frac{1}{2}\gamma_{\mu}\gamma^{(5)} \theta_+ \gamma^m
\partial_m(D-\phi)$$ and inserting the latter expression in (\ref{4dgrav}) we obtain
\begin{equation}
\label{etap}
f W e^{-(D-\phi)}\eta_++\frac{f^*}{2} \partial_m(D-\phi) \gamma^m\eta_--
\frac{f^*}{12}\tilde\Sigma_{mnp}\gamma^{mnp}\eta_-=0.
\end{equation}
Multiplying now eq.~(\ref{etap}) with $\eta^\dagger_+$ from the left
we obtain
\begin{equation}
\label{gi1}
f We^{-(D-\phi)} -\frac{f^*}{12}\tilde\Sigma_{mnp}(\eta^\dagger_+
\gamma^{mnp}\eta_-)=0 \;\Longrightarrow\; f We^{-(D-\phi)}-\frac{f^*}{12}
(\tilde \Lambda^*)^3\|\Omega^*\|^2=0,
\end{equation}
where $\|\Omega\|=\Big(\Omega^*_{mnp}\Omega^{mnp}\Big)^{1/2}$. Using
Fierz identities one can find that $\|\Omega\|^2=48$. For vanishing
total condensate $\tilde\Lambda^3=0$, we obtain $W=0$ and hence
$M_{1,3}$ is flat. However in general it holds that
\begin{equation}
\hat{R} \;\sim\;  e^{2(D-\phi)} |\tilde\Lambda|^6 .
\end{equation}

Now we can  prove the constancy of $D-\phi$. By multiplying
eq.~(\ref{etap}) with $\eta^\dagger_- \gamma_n$ from the left and
using the equality $\eta^\dagger_-\gamma^{mnpq}\eta_-=-3 J^{[mn}
J^{pq]}$ we obtain
\begin{equation}
f^* \partial_n(D-\phi)-\frac{i}{2} f^*\partial_m(D-\phi) J_n^{\;\;m}-
\frac{3 }{256} f^* \tilde\Sigma^{mpq}J^{[mn} J^{pq]}
+\frac{3i}{256}f^*
\tilde\Sigma_{nmp}J^{mp}=0.
\end{equation}
Since $\tilde\Sigma$ is a $(3,0)+(0,3)$ form, we have
$\tilde\Sigma_{nmp}J^{mp}=0$ and $\tilde\Sigma_{mpq} J^{[nm} J^{pq]}=0$. This immediately
shows
that
\begin{equation}
\label{constantdphi}
\partial_n (D-\phi)=0,
\end{equation}
and demonstrates that the constancy of $D-\phi$ is actually imposed
upon us from the supersymmetry conditions.

\subsection{Conditions on the internal geometry and intrinsic torsion classes}

\subsubsection{Gravitino variation}

Next we do the analysis of the 6-dimensional part of the gravitino
variation. This reads as
\begin{equation}\label{3.18}
\delta\psi_m=\nabla_m\epsilon-\frac{1}{4}\Big(\hat H_{mpq}-
2\tilde\Sigma_{mpq}+\frac{1}{6}\Delta_{mpq}\Big)
\Gamma^{pq}\epsilon-\frac{1}{12}\tilde\Sigma_{npq}\Gamma_m
\Gamma^{npq}\epsilon=0.
\end{equation}
Inserting the spinor ansatz (\ref{spinans}) in eq.~(\ref{3.18}) we
obtain
\begin{equation}
\label{gve}
 \nabla_m\eta_+=-\partial_m \log f \eta_++\frac{1}{4}\Big(\hat H_{mpq}-
2\tilde\Sigma_{mpq}+\frac{1}{6}\Delta_{mpq}\Big)
\gamma^{pq}\eta_++\frac{1}{12}\tilde\Sigma_{npq}\gamma_m\gamma^{npq}\eta_+
.
\end{equation}
Using the identity
$$\gamma^m\gamma^{npq}=\gamma^{mnpq}+\delta^{mq}\gamma^{np}+\delta^{mp}\gamma^{qn}+
\delta^{mn}\gamma^{pq}$$ we can rewrite eq.~(\ref{gve}) as
\begin{equation}
\nabla_m\eta_+=-\partial_m \log f
\eta_++\frac{1}{4}\Big(\hat H_{mpq}-\tilde\Sigma_{mpq}+\frac{1}{6}\Delta_{mpq}\Big)
\gamma^{pq}\eta_++\frac{1}{12}\tilde\Sigma_{npq}\gamma_{m}^{\ \ npq}\eta_+
.
\end{equation}
Expressing $\gamma_{m}^{\ \ npq}$ as $\gamma_{m}^{\ \ npq}= -
\frac{i}{2} \epsilon_{m}^{\ \ npqrs} \gamma_{rs} \gamma_{7}$ we then
obtain
\begin{equation}\label{nablaeta}
\nabla_m\eta_+=-\partial_m \log f \eta_++\frac{1}{4}\Big( \hat
H_{mpq}- \tilde\Sigma_{mpq}+ \frac{1}{6}\Delta_{mpq}+ i
\ast\!\tilde\Sigma_{mpq}\Big) \gamma^{pq}\eta_+,
\end{equation}
where $ \ast\tilde\Sigma_{mpq} =
\frac{1}{3!}\tilde\Sigma_{klr}\epsilon^{klr}_{\ \ \   mpq}$.

However, the total contribution of $\tilde\Sigma_{mnp}$ in the above
equations drops out. Indeed, due to the imaginary
(anti)-self-duality of $\Omega_{mnp}$ ($\Omega^*_{mnp}$) we have
$\ast\tilde\Sigma_{mpq}=-i \tilde\Lambda^3 \Omega_{mpq} + c.c.$.
Using also the fact that $\Omega_{mpq} \gamma^{pq}\eta^+=0$,   we
finally obtain
\begin{equation}\label{nablaetalast}
\nabla_m\eta_+=-\partial_m \log f \eta_++\frac{1}{4}\widehat H_{mpq}
\gamma^{pq} \eta_+ ,
\end{equation}
where we have defined $\widehat H_{mnp}=\hat
H_{mnp}+\frac{1}{6}\Delta_{mnp}$. Consequently, the spinor $\eta^+$
is parallel under the connection with torsion
$\nabla^{(-)}=\nabla-\frac{1}{4} \widehat H$. Defining the curvature
through
$[\nabla^{(-)}_m,\nabla^{(-)}_n]\eta^+=\frac{1}{4}R^{(-)}_{mnpq}
\gamma^{pq} \eta^+$, the integrability condition for
eq.~(\ref{nablaetalast}) implies
\begin{equation}
R^{(-)}_{mnpq} J^{pq}=0.
\end{equation}
Now using  eq.~(\ref{nablaetalast}) and its conjugate we find that $
|f| = const$. Hence, in general $f=e^{i\alpha}$. For our purposes,
however, taking $f=1$ is not restrictive  and we will do so.

It is straightforward to compute the covariant derivatives of the
$SU(3)$-structure forms:
\begin{eqnarray}
\nabla_m J_n^{\;\;p}&=&
\Big(\widehat H_{sm}^{\;\;\;\;p}J_n^{\;\;s}+
\widehat H^s_{\;\;mn}J_s^{\;\;p}\Big),
\label{delj}\\
\nabla_m \Omega_{kln}&=& \widehat H^p_{\;\;mn} \Omega_{pkl}+\widehat H^p_{\;\;lm} \Omega_{pkn}
+\widehat H^p_{\;\;mk} \Omega_{pln}\label{delo}.
\end{eqnarray}
These relations imply that the internal space has contorsion given
by $\widehat H_{mnp}$, which in turn has a contribution due to the
H-flux and another due to the dilatino condensate.

\subsubsection{Dilatino variation}

In the case that a non-zero gaugino condensate is present, the
vanishing of the dilatino variation demands that $\hat H_{mnp}$ has
a non-zero $(3,0)+(0,3)$ piece. Indeed, the dilatino variation is
\begin{equation}
\delta\lambda=\Gamma^m\nabla_m\phi
\epsilon+\frac{1}{24}\Big(\hat H_{mnp}+
\Sigma_{mnp}\Big)\Gamma^{mnp}\epsilon=0,
\end{equation}
and for the spinor ansatz (\ref{spinans}) it becomes
\begin{equation}
\label{dil}
\partial_m\phi\gamma^m \eta_++\frac{1}{24}\Big(\hat H_{mnp}+\Sigma_{mnp}\Big)\gamma^{mnp}
\eta_+=0.
\end{equation}
Multiplying eq.~(\ref{dil}) by $\eta^\dagger_-$ from the left gives
\begin{equation}
\Big(\hat H_{mnp}+\Sigma_{mnp}\Big) \Omega^{mnp}=0.
\end{equation}
This fixes the $(3,0)$ and $(0,3)$ parts of the supercovariantized
flux in terms of the
gaugino condensate vev
\begin{equation}\label{hath3.0}
\hat H^{(3,0)}_{mnp}=\Lambda^3 \Omega_{mnp}.
\end{equation}
If the dilatino condensate is zero, so that $\hat
H_{mnp}=H_{mnp}$, the above equations imply that in order to
preserve supersymmetry in the presence of a non-vanishing gaugino
condensate an H-flux with non-vanishing  $(3,0)$ and $(0,3)$
components should be present too.

Multiplying now eq.~(\ref{dil}) by $\eta^\dagger_+ \gamma_k$ from
the left gives
\begin{equation}\label{3.30}
i J_k^{\;\;n} \partial_n\phi+\partial_k \phi-\frac{1}{8}\hat H_{mnp} J_k^{\;\;m}J^{np}+
\frac{i}{8}\hat H^{kmn}J^{mn}=0.
\end{equation}
The contribution of $\Sigma_{mnp}$ drops out since the
gaugino condensate has only $(3,0)+(0,3)$ components. From
eq.~(\ref{3.30}) we obtain an equation for the dependance of the
dilaton on the internal space coordinates
\begin{equation}
\partial_k \phi=\frac{1}{8}\hat H_{mnp} J_k^{\;\;m}J^{np}=\frac{i}{8}( H^{(1,0)}
- H^{(0,1)}).
\label{pphi}
\end{equation}

We see that only the non-primitive $(2,1)+(1,2)$ piece of $\hat
H_{mnp}$ (and hence of $ H_{mnp}$ since they only differ in their
$(3,0)+(0,3)$ parts), contributes to the variation of the dilaton.
In particular, if we have only non-zero the $(3,0)$ and $(0,3)$
components of the H-flux, the dilaton is constant, $\partial_k \phi
=0$. Then due to eq.~(\ref{constantdphi}) the warp factor $D(y)$ is
constant too. This shows that the assumptions of
refs.~\cite{Govindarajan:1986iz, Govindarajan:1986kb} on the
constancy of $D$ and $\phi$ are not just sufficient but also
necessary for unbroken supersymmetry.

\subsubsection{Gaugino variation}

Finally we consider the gaugino variation. This is given by
\begin{equation}
\label{gv} \delta\chi=-\frac{1}{4} \hat F_{mn}\Gamma^{mn} \epsilon +
\frac{1}{32} \Big(3
\Phi-\frac{3}{2}\Phi_{mn}\Gamma^{mn}-\frac{1}{24}\Phi_{mnkl}
\Gamma^{mnkl}\Big)\epsilon,
\end{equation}
while inserting  the spinor ansatz and the decomposition  of the
condensates yields
\begin{equation}
\label{gvv} \delta\chi=-\frac{1}{4}(F_{mn}+\frac{3}{16}\Phi_0
J_{mn})\gamma^{mn} \eta_+,
\end{equation}
where the terms proportional to $\Phi$ are zero due to the identity
$(1+ \frac{1}{24} J_{mn} J_{kl}
\gamma^{mnkl})\eta_+=0$.

A vanishing gaugino variation demands that the field strength
satisfies $F_{mn}=J_m^{\ k} J_n^{\ l} F_{kl}$,
i.e.~$F^{(2,0)}=F^{(0,2)}=0$. Furthermore, the
non-primitive part of $F_{mn}$ is compensated by the
vev of the dilatino-gaugino condensate $\Phi_0$ as
$\Phi_0=-\frac{8}{9} F_{mn} J^{mn}$.
In the usual
case where this condensate is taken to be zero, we end up with
the standard result that a supersymmetry preserving background
gauge field has to be a primitive $(1,1)$ form.

\subsection{Torsion classes}

We can summarize now our findings in the language of intrinsic torsion classes
which are defined as
\begin{equation}
dJ=\frac{3i}{4}({\cal W}_1 \Omega^*-\bar {\cal W}_1 \Omega)+{\cal
W}_4 \wedge J+ {\cal W}_3,
\end{equation}
and
\begin{equation}
d\Omega = {\cal W}_1 J \wedge J + {\cal W}_2 \wedge J + {\cal W}^*_5
\wedge\Omega,
\end{equation}
satisfying $J\wedge {\cal W}_3=J\wedge J\wedge {\cal W}_2=\Omega\wedge {\cal W}_3=0$.
The classes ${\cal W}_1$ and ${\cal W}_2$ can be decomposed
in real and imaginary parts as ${\cal W}_1=
{\cal W}_1^++{\cal W}_1^-$ and ${\cal W}_2=
{\cal W}_2^++{\cal W}_2^-.$

The classes ${\cal W}_1$ and ${\cal W}_2$ are vanishing when the
almost complex structure is integrable, i.e.~when the manifold is
complex. A K\"ahler manifold has furthermore ${\cal W}_3={\cal
W}_4=0$. Finally, CY manifolds have in addition ${\cal W}_5=0$.
Hence, one can think of the five torsion classes as parameterizing
the deformation away from $SU(3)$-holonomy.

Using eqs.~(\ref{delj}) and (\ref{delo}) we obtain
\begin{eqnarray}
{\cal W}_1 &=& \frac{1}{6}\widehat H^{(0,3)} =
8 \Big((\Lambda^*)^3-\frac{1}{6}(\delta^*)^3\Big), \nonumber\\
{\cal W}_2 &=& 0, \nonumber\\
{\cal W}_3&=&i(H^{o(2,1)}-H^{o(1,2)}),\label{torsion}\\
{\cal W}_4&=&-\frac{i}{2}(H^{(1,0)}-H^{(0,1)}),\nonumber\\ {\cal W}_5&=&
i(H^{(1,0)}-H^{(0,1)})\nonumber.
\end{eqnarray}
In order to derive the class ${\cal W}_1$, we took into account
eq.~(\ref{hath3.0}) and the definition $\widehat H_{mnp}=\hat
H_{mnp}+\frac{1}{6}\Delta_{mnp}$. Using eq.~(\ref{pphi}) we can
express the torsion classes ${\cal W}_4$ and ${\cal W}_5$ in terms
of the dilaton gradient as
\begin{equation}
{\cal W}_4=-4 d\phi, \;\;\; {\cal W}_5=8 d\phi.
\end{equation}

Now we summarize our findings in order to be easy to contrast them
to the case with H-flux but without condensates
\cite{Cardoso:2002hd} and to the case with H-flux and only the gaugino
condensate \cite{Frey:2005zz}. We have found that:
\begin{itemize}
\item The gaugino condensate $\Sigma_{mnp}$ induces a $(3,0)+(0,3)$ piece
to the supercovariant $\hat H$-flux.
\item The dilatino and gaugino condensate yield a non-vanishing ${\cal W}_1$
class, rendering the internal spaces non-complex.
\item The class ${\cal W}_2$ is zero for the H-flux and for all
the condensates we turned on.
\item The intrinsic torsion classes ${\cal W}_i, \; i=3,4,5$
are determined only in terms of
the $(2,1)+(1,2)$ pieces of the flux and hence do not depend on
the condensates.
\item The spacetime curvature depends on the dilatino and gaugino condensates
in such a way that one can tune them to obtain a Minkowski vacuum with
a non-complex internal space \footnote{A similar conclusion was
reached in ref.~\cite{Duff:1982yi} for compactifications of
11-dimensional supergravity with gravitino condensates.}.

\item The dilatino-gaugino condensates in the gaugino variation allow for background
gauge fields that can have a non-primitive $(1,1)$ piece.
\end{itemize}

\section{Supersymmetric solutions on nearly-K\"ahler spaces}

In the previous section we presented a set of conditions on the
intrinsic torsion classes of a six-dimensional manifold with
$SU(3)$-structure, which are necessary for obtaining supersymmetric
vacua of heterotic string theory in the presence of H-flux and
several condensates. The most general manifolds with torsion classes
specified by eq.~(\ref{torsion}) are known as ${\cal G}_1$ manifolds
\cite{grey}.

In the absence of condensates, only an H-flux with non-trivial
$(2,1)+(1,2)$ components can be present if some  supersymmetry is to
be preserved. In particular, if only the primitive part of the flux
is non-zero we have ${\cal W}_1={\cal W}_2={\cal W}_4={\cal W}_5=0$.
These manifolds are known as special-hermitian and are well-studied
in the mathematical literature. However, so far it has been proved that is
difficult to satisfy the Bianchi identity for
compactifications of this type.

Another natural choice is to assume that only the condensates and
the corresponding pieces of the H-flux induced by them are
non-vanishing. In this case, the H-flux is of  $(3,0)+(0,3)$ type
and, according to eq.~(\ref{torsion}), the appropriate
six-dimensional manifolds will have only the class ${\cal W}_1$
different from zero. Notice that supersymmetry specifies further the
H-flux in terms of the condensates, effectively determining  ${\cal
W}_1$ in terms of $\Lambda^3$ and $\delta^3$. Setups with
$\delta^3=0$ have been already considered in
refs.~\cite{Govindarajan:1986iz, Govindarajan:1986kb}. One of our
purposes here is to revisit them in the framework of
$SU(3)$-structures and intrinsic torsion. Besides that, the analysis
of the previous section showed that some of the assumptions of
refs.~\cite{Govindarajan:1986iz, Govindarajan:1986kb} are actually
necessary and not just sufficient for preserving spacetime
supersymmetry. Furthermore, the authors of
refs.~\cite{Govindarajan:1986iz, Govindarajan:1986kb} did not take
into account the modifications to the hererotic Bianchi identities
due to the non-trivial torsion. As we shall see,  to satisfy the
Bianchi identity, the presence of the gaugino condensate is not
enough.

Six-dimensional manifolds with non-zero ${\cal W}_1$ have been
well-studied too in the mathematical literature. They are known as
nearly-K\"ahler and they have certain special properties. Among
others, they are Einstein spaces of positive scalar curvature, their
almost complex structure is never integrable since the Nijenhuis
tensor is non-zero, their first Chern class vanishes, and they admit
a spin structure. In the ensuing we will discuss in detail some
examples of nearly-K\"ahler spaces which are based on coset spaces.

Let us just mention for the moment that it would be interesting to
relax the condition $H^{(2,1)}=0$ and turn-on a primitive
$(2,1)+(1,2)$ piece of the H-flux (in addition to the $(0,3)+(3,0)$
piece induced by the condensates). Then the conditions on the
intrinsic torsion would imply that the compactification manifold
should have ${\cal W}_2={\cal W}_4={\cal W}_5=0$ and (choosing
purely imaginary condensates) ${\cal W}_1^+=0$. Such manifolds are
particular cases of half-flat manifolds and one simple but
interesting realization is provided by twisted toroidal orbifolds
\cite{Camara:2005dc} (see also \cite{Dall'Agata:2005fm} for the
7-dimensional analogue). It would be worthwhile to investigate if
more general solutions can be provided by such spaces, but we
postpone that for future work.

\subsection{Nearly-K\"ahler coset spaces}

The only known examples of compact nearly-K\"ahler spaces are
3-symmetric spaces that can be described as homogeneous cosets.
These are {\bf(i)} $G_{2}/SU(3)$, which is an $S^6$ but less
symmetric than the usual 6-sphere, {\bf(ii)} $ Sp(4)/(SU(2)\times
U(1))_{non-max.}$ which is similarly a less symmetric version of
$\mathbb{CP}_3$, {\bf(iii)} $ SU(3)/(U(1)\times U(1))$ which is the
flag manifold $F(1,2)$, and  {\bf(iv)}$SU(2)^3/SU(2)$ which is
isomorphic to $S^3 \times S^3$. All these spaces admit an
$SU(3)$-structure. Although in general they are half-flat, there are
special values of their moduli for which only ${\cal W}_1^-$ is
non-zero.  For these special values these cosets become nearly-K\"ahler.

In Appendix B we present the $SU(3)$-structures of the 4 cosets and
compute their intrinsic torsion classes. Inspection of the results
shows that all possible radii of the cosets have to be the same if
we want to keep a vanishing ${\cal W}_2$ class. Then, the only
non-zero class is ${\cal W}_1$ and  it takes the general form
${\cal W}_1=-i \frac{w }{\sqrt{a}}$. The actual values of $w$ for
each coset can be found in Appendix B.

Furthermore,
using the corresponding
curvature 2-forms \cite{Mueller-Hoissen:1987cq}, we can compute the first
Pontrjagin classes:
\begin{eqnarray}
G_{2}/SU(3)&:& {\rm tr}  R \wedge  R = 0,\nonumber\\
Sp(4)/(SU(2)\times U(1))&:& {\rm tr}  R \wedge  R = \frac{18}{a^2} J \wedge J,\label{rr}\\
SU(3)/(U(1)\times U(1))&:& {\rm tr}  R \wedge  R = \frac{18}{a^2}
J \wedge J,\nonumber\\
SU(2)^3/SU(2) &:& {\rm tr}  R \wedge  R = -(\frac{8}{9a})^2 J \wedge J.\nonumber
\end{eqnarray}
Hence, for all cosets under consideration and for metrics with
nearly-K\"ahler structure we have a general formula for the
first Pontrjagin class given by ${\rm tr}  R \wedge  R = \frac{p_1}{a^2}
J\wedge J$. Since $J\wedge J$ is exact, this class is
cohomologically trivial. Using the general expression for ${\cal
W}_1$ we can also write
\begin{equation}
\label{p1}
{\rm tr}  R \wedge  R = \frac{p_1}{w^4} |{\cal W}_1|^4 J \wedge J.
\end{equation}

The supersymmetry conditions demand that the torsion class ${\cal
W}_1$ is fixed in terms of the condensates.  This stabilizes the
radial modulus to a value
\begin{equation}\label{tobe}
a=- \frac{w^2}{8} \Big|\Lambda^3-\frac{1}{6}\delta^3\Big|^{-2}.
\end{equation}
Since ${\cal W}_1$ is imaginary for the cosets under consideration,
the condensate combination $(\Lambda^*)^3-\frac{1}{6}(\delta^*)^3$
has to be imaginary too.

\subsection{Bianchi identity}

Let us now consider the Bianchi identity for the H-flux. The full
Bianchi identity includes a correction due to the gravitational
Chern-Simons term required for anomaly cancelation
\cite{Green:1984sg} and it reads
\begin{equation}\label{BId}
dH=\frac{1}{2} \Big({\rm tr}  R^{(+)} \wedge  R^{(+)} - \frac{1}{30}
{\rm Tr} F\wedge F\Big),
\end{equation}
where $R^{(+)}$ is the curvature 2-form of the connection
$\nabla^{(+)} = \nabla + \frac{1}{4} \widehat H$. 
We have also adopted
the usual normalization for the trace of the gauge field strengths.

The torsion of the connection whose curvature appears in the Bianchi
identity is the opposite of the torsion of the connection
$\nabla^{(-)} = \nabla - \frac{1}{4}\widehat H$  appearing in the
supersymmetry variation of the gravitino in the internal space. Although this fact
is well-established only for the case where the torsion comes entirely
from the H-flux
\cite{Hull:1986kz}, it is quite natural to expect that the proper
generalization of the gravitational Chern-Simons correction
to the Bianchi involves the full torsion tensor appearing in the gravitino
variation.
Let us also mention that demanding only supersymmetry and anomaly freedom
does not specify completely the torsion relevant for
the gravitational Chern-Simons term
\cite{Hull:1985dx}. From this point of view our choice is
perfectly consistent. Notice, furthermore, that the ambiguity is fixed by the 
additional requirement
of conformal invariance on the worldsheet \cite{Hull:1986kz} or, equivalently, of
satisfying the spacetime equations of motion. It would be
extremely interesting and important to actually verify the validity of our
choice with an explicit calculation. However,
such a computation is bound to be subtle due to the presence
of the condensates.

We can now proceed with the computation of the quantities
that appear in the Bianchi identity.
The curvature tensor for $\nabla^{(+)} = \nabla +
\frac{1}{4}\widehat H$ is
\begin{equation}
 R^{(+)}_{mnpq}=R_{mnpq}+2 \nabla_{[m} \widehat H_{n]pq} -
(\widehat H_{mp}^{\;\;\;\;\;r}\widehat H_{rnq}- \widehat
H_{np}^{\;\;\;\;\;r}\widehat H_{rmq}),
\end{equation}
and the corresponding curvature 2-form is
\begin{equation}
R^{(+)}_{mn} = \frac{1}{2}R^{(+)}_{mnpq}e^{p} \wedge e^{q}.
\end{equation}

For nearly-K\"ahler spaces the torsion $\widehat H_{mnp}$ has only
$(3,0)+(0,3)$ parts which are fixed by the class ${\cal W}_1$ as
\begin{equation}
\widehat H_{mpq}=8 \ ({\cal W}_1 \Omega_{mpq}+c.c.).
\end{equation}

Using eq.~(\ref{delo}) and the identity $$\Omega_{pmn} \Omega^{*pkl}
= 16 (\Pi^+)_{[m}^{\ \ [k} (\Pi^+)_{n]}^{\ \ l]}$$ (c.f.~for
instance \cite{Lust:2004ig}), we obtain
\begin{eqnarray}
\nabla_{[m} \widehat H_{n]pq}&=&24 |{\cal W}_1|^2 J_{[mn}J_{pq]},\label{dh}\\
\widehat H_{mp}^{\;\;\;\;\;r}\widehat H_{rnq}-\widehat
H_{np}^{\;\;\;\;\;r}\widehat H_{rmq}&=&4 |{\cal W}_1|^2
\Big((g_{mp}g_{nq}-g_{mq}g_{np})-J_{mn}J_{pq}-3J_{[mn}J_{pq]}\Big).
\end{eqnarray}
Subsequently, the generalized curvatures are found to be
\begin{eqnarray}
\label{curvplus} R^{(+)}_{mnpq}&=&R_{mnpq}+60 |{\cal W}_1|^2
J_{[mn}J_{pq]}
-4 |{\cal W}_1|^2  \Big((g_{mp}g_{nq}-g_{mq}g_{np})-J_{mn}J_{pq}\Big),\\
R^{(-)}_{mnpq}&=&R_{mnpq}-36 |{\cal W}_1|^2  J_{[mn}J_{pq]} -4
|{\cal W}_1|^2   \Big((g_{mp}g_{nq}-g_{mq}g_{np})-J_{mn}J_{pq}\Big).
\end{eqnarray}

The left-hand-side of the Bianchi identity (\ref{BId}) can be
derived by using the fact that the $\hat H_{mnp}\equiv H_{mnp}-6
\Delta_{mnp}$ is fixed, by the vanishing of the dilatino variation,
to be $\hat H_{mnp}=\ \Lambda^3 \Omega_{mnp}+c.c.$. Since for
nearly-K\"ahler spaces holds that $d\Omega = {\cal W}_1 J \wedge J$,
we find that
\begin{equation}
dH= \  (\Lambda^3-6\delta^3){\cal W}_1 J\wedge J + c.c.
\end{equation}
Then we can eliminate the explicit dependance on $\Lambda$ and its
conjugate by using the supersymmetry condition ${\cal W}_1= 8 \
\Big((\Lambda^*)^3-\frac{1}{6}(\delta^*)^3\Big)$. Finally we obtain
\begin{equation}\label{4.13}
dH=\Big(\frac{1}{4} |{\cal W}_1|^2 - {\rm Re}(\delta^3 {\cal
W}_1)\Big) J\wedge J ,
\end{equation}
where we have rescaled $\delta^3 \rightarrow \frac{18}{35}\delta^3$.

The right-hand side of the Bianchi identity (\ref{BId}) depends on
the first Pontrjagin class of $\nabla^{(+)}$. After a slightly tedious
calculation we obtain
\begin{equation}
{\rm tr} R^{(+)} \wedge R^{(+)} = {\rm tr} R\wedge R - 4608 |{\cal
W}_1|^4  J\wedge J.
\end{equation}
Next,  to obtain a feeling of the implications of the Bianchi identity
(\ref{4.13}), we assume that there is no background gauge field.
Using eq.~(\ref{p1})
 we find that
the Bianchi identity reduces to the following equation
\begin{equation}
\Big(\frac{1}{4} |{\cal W}_1|^2 -\  {\rm Re}(\delta^3 {\cal
W}_1)\Big)= \  \Big(\frac{p_1}{2w^4} -2304\Big) |{\cal W}_1|^4.
\end{equation}
In addition, for simplicity let us consider the case $G_2/SU(3)$ for
which $p_1=0$. Recall that for all cosets the class ${\cal W}_1$ is
imaginary. Then we find the following solution for  $\delta^3$,
which in the present case is purely imaginary:
\begin{equation}
\ \delta^3=-i|{\cal W}_1|\Big(\frac{1}{4}+2304 \  |{\cal
W}_1|^2\Big).
\end{equation}
Moreover, the gaugino condensate is fixed to a purely imaginary
value as well given by
\begin{equation}
\ {\rm Im}(\Lambda^3)=\frac{3}{35}{\rm Im}(\delta^3)-\frac{1}{8}
|{\cal W}_1|,
\end{equation}
which in terms of the class ${\cal W}_1$ becomes
\begin{equation}
\ {\rm Im}(\Lambda^3)=-\frac{1}{35}
|{\cal W}_1|\Big(\frac{41}{8}+3 \cdot 2304 \  |{\cal W}_1|^2\Big).
\end{equation}
It is straightforward to find similar solutions for the other three
cosets.

It is clear that the previous solution retains in four dimensions
the full gauge group $E_8 \times E_8$  and thus it is undesirable as
a realistic GUT. However, we can  turn on  a background gauge field
that breaks part of the gauge symmetry in a supersymmetric way. The
condition for supersymmetry dictates that the gaugino variation
(\ref{gvv}) is zero. A simple way to deal with this is to assume
that all the condensates $\Phi_{[\cdots]}$ are vanishing and to
consider further the curvature 2-form corresponding to the $SU(3)$
connection with torsion $\nabla^{(-)}$. Embedding this connection in
the gauge group, yields a background $SU(3)$ gauge field $ (F_{ab})_
{mn} =R^{(-)}_{ab mn}$ which leads to a vanishing gaugino variation
as a consequence of the integrability condition $R^{(-)}_{abmn}
\gamma^{mn} \eta^+=0$. For such a choice $\frac{1}{30} {\rm Tr}
F\wedge F = {\rm tr} R^{(-)} \wedge R^{(-)}$, where the latter is
found to be
\begin{equation}
{\rm tr} R^{(-)} \wedge R^{(-)} = {\rm tr} R\wedge R + 1536 |{\cal W}_1|^4
J\wedge J.
\end{equation}
The Bianchi identity leads now to the relation
\begin{equation}
\Big(\frac{1}{4} |{\cal W}_1|^2 -\  {\rm Re}(\delta^3 {\cal
W}_1)\Big)= -3072 |{\cal W}_1|^4,
\end{equation}
which fixes the dilatino condensate and, in turn, also the gaugino
condensate by the supersymmetry condition (\ref{tobe}).
 The background gauge field breaks the gauge group down to $E_{6}
\times E_{8}$ as does the standard embedding in the case of CY
compactifications.

Using now the fact that $\Omega_{mnp}$ ($\Omega_{mnp}^*$) is imaginary
(anti)-self-dual, one can easily check that for all the above solutions
the NS-NS flux is coclosed:
\begin{equation}
d*H=0
\end{equation}
and hence it solves the source-free equation of motion.
Accordingly, since the Bianchi equation and the equation of motion for $H$
are satisfied and the background preserves some supersymmetry,
the Einstein equations are also verified
as a consequence of the results of \cite{Gauntlett:2002fz}.

\section{Conclusions}
In the present work we have examined in detail some examples of
supersymmetric AdS compactifications of heterotic strings on
non-complex manifolds, which  we believe will be eventually
phenomenologically interesting. The considered six-dimensional
manifolds include the coset spaces $G_{2}/SU(3)$,
$Sp(4)/(SU(2)\times U(1))_{non-max}$, $SU(3)/(U(1)\times U(1))$ and
$SU(2)^{3}/SU(2)$. These have been examined some time ago in the
context of Coset Space Dimensional Reduction
\cite{Forgacs:1979zs,Kubyshin:1989vd,Kapetanakis:1992hf}, in
attempts to obtain realistic GUTs from extra dimensions
\cite{Lust:1985be,Manousselis:2004xd,Manousselis:2001re,Bais:1985yd,Kapetanakis:1990rz},
and also in the context of heterotic sting theory in studies of
possible supersymmetry preserving backgrounds in refs.~\cite{
Castellani:1986rg, Lust:1986ix, Govindarajan:1986iz,
Govindarajan:1986kb}. These solutions have been analyzed here in the
framework of $G$-structures while a new example based on the coset
$SU(2)^3/SU(2)$ has been added.

The conjecture that nearly-K\"ahler spaces have no complex-structure
moduli \cite{Micu:2004tz} and the fact that the supersymmetry
preserving conditions fix the radii to specific values, implies that
one should expect supersymmetric AdS vacua with most moduli
stabilized in this approach. Subsequently,  it would be
worthwhile to analyze the possibilities for "lifting" the AdS vacuum
to Minkowski or de Sitter \cite{Kachru:2003aw, Burgess:2003ic} and
study supersymmetry breaking in this context. Work in this direction
is currently in progress.

\vspace{2cm}

\noindent
{\bf Acknowledgments}

\medskip

We would like to thank K.~Behrndt, G.L.~Cardoso, G.~Dall'Agata,
J.-P.~Derendinger, A.~Frey, P.~Forgacs, J.~Gauntlett, M.~Lippert,
 D.~L\"ust, A.~Micu, A.~Nagy,
G.~Papadopoulos, K.~Sfetsos, A.~Uranga and D.~Zoakos for helpful discussions and
correspondence. We are also grateful to  A.~Frey, 
M.~Lippert and D.~Tsimpis for interesting comments and questions  
on the first version of this paper, as well as
to T.~Kimura for pointing out a numerical mistake.

The work is supported by the EPEAEK programme ``Pythagoras" and
co-founded by the European Union (75\%) and the Hellenic state
(25\%). The work of N.P. is supported by the Swiss National Science
Foundation and by the Commission of the European Communities under
contract MRTN-CT-2004-005104.
\medskip

\section*{Appendix A: Gamma matrix conventions}

The 10-dimensional Gamma matrices $\Gamma^M$ with $M=0,1,\ldots,9$
satisfy $\{\Gamma^M,\Gamma^N\}=2\eta^{MN}$ where $\eta^{MN}=(-1,1,\ldots,1)$.
Splitting to 4+6 indices $\mu,\nu=0,1,2,3$ and $m,n=4,\ldots,9$ we can write
them as $\Gamma^\mu=\gamma^\mu\otimes\mathbb{I}$ and $\Gamma^m=\gamma^5\otimes\gamma^m$
where $\gamma^\mu$ the usual 4-dimensional gamma matrices, $\gamma^{(5)}$ is  the
chirality matrix satisfying $(\gamma^{(5)})^2=1$ and $\{\gamma^\mu,\gamma^{(5)}\}=0$,
and $\gamma^m$ are the Gamma matrices in 6-dimensional Euclidean space.

The 6-dimensional spinors $\eta_\pm$ have opposite chiralities
$\gamma_{(7)} \eta_\pm=\pm\eta_\pm$ where $\gamma_{(7)}=-i \gamma_4 \ldots\gamma_9$.
They satisfy $\eta^\dagger_\pm=\eta^T_\mp C_{(6)}$ where $C_{(6)}$ is the 6-dimensional
charge conjugation matrix. The 4-dimensional Weyl spinors $\theta_\pm$
are also chiral $\gamma_{(5)} \theta_\pm=\pm\theta_\pm$ with $\gamma_{(5)}=i
\gamma_0\gamma_1\gamma_2\gamma_3$. Moreover $\bar\theta_\pm=\pm \theta^T_\mp C_{(4)}$
 with $C_{(4)}$ the 4-dimensional charge conjugation matrix. As usual,
 for a Dirac spinor $\psi$ we define
 $\bar\psi=\psi^\dagger\gamma^0$.

The 10-dimensional chirality and charge conjugation matrix are
$\Gamma_{(11)}=\gamma_{(5)}\otimes\gamma_{(7)}$ and
$C_{(10)}=C_{(4)}\gamma_{(5)}\otimes C_{(6)}$. With these
definitions the spinor (\ref{spinans}) satisfies
$\Gamma_{(11)}\epsilon=\epsilon$ and
$\bar\epsilon=\epsilon^TC_{(10)}$, i.e.~it is chiral and Majorana.
For the gamma matrix manipulations we used \cite{Gran:2001yh}.

Finally we present some spinor bilinears that were used in the
considerations of section 3:

\begin{eqnarray}
\eta_+^\dagger\eta_+&=&\eta_-^\dagger\eta_-=1,\\
\eta_-^\dagger\eta_+&=&\eta_+^\dagger\eta_-=0,\\
\eta^\dagger_\pm\gamma^m\eta_\pm&=&\eta^\dagger_\mp\gamma^m\eta_\pm=0,\\
\eta^\dagger_\pm\gamma^{mn}\eta_\mp&=&\eta^\dagger_\pm\gamma^{mnpq}\eta_\mp=0,\\
\eta^\dagger_\pm\gamma^{mn}\eta_\pm&=&\pm i J^{mn},\\
\eta^\dagger_\pm\gamma^{mnpq}\eta_\pm&=&-3 J^{[mn} J^{pq]},\\
\eta^\dagger_\pm\gamma^{mnp}\eta_\pm&=&\eta^\dagger_\mp\gamma^{mnpqrs}\eta_\pm=0,\\
\eta^\dagger_\pm\gamma^{mnpqr}\eta_\pm&=&\eta^\dagger_\mp\gamma^{mnpqr}\eta_\pm=0.
\end{eqnarray}

\section*{Appendix B: $SU(3)$-structures on coset spaces}

In \cite{House:2005yc} the intrinsic torsion classes for the coset
$SU(3)/(U(1)\times U(1))$ were calculated in terms of the radii $a,b,c$.
Here we extend this computation to the rest of the coset
spaces that admit a nearly-K\"ahler structure.
The data required for this exercise can be found in \cite{Mueller-Hoissen:1987cq,
Gutowski:2002bc}.

\subsection*{$SU(3)/(U(1)\times U(1))$}

$\bullet$ Metric:
\begin{equation}
ds^2=a (e^1 \otimes e^1 + e^2 \otimes e^2) + b (e^3 \otimes e^3 +
e^4 \otimes e^4) + c (e^5 \otimes e^5 + e^6 \otimes e^6).
\end{equation}
$\bullet$ $SU(3)$-structure:
\begin{eqnarray}
J &=& -a e^{1} \wedge e^{2} + b e^{3} \wedge e^{4} - c e^{5} \wedge
e^{6}, \\
dJ &=& -(a + b + c) (e^{1} \wedge e^{3} \wedge e^{5} +
        e^{1} \wedge e^{4} \wedge e^{6}  - e^{2} \wedge e^{3} \wedge e^{6} +
        e^{2} \wedge e^{4} \wedge e^{5}), \\
\Omega &=& \sqrt{( a b c)}(e^{1} + i e^{2}) \wedge (e^{3} - i e^{4})
\wedge (e^{5} + i e^{6}),\\
d\Omega &=& 4i \sqrt{abc} ( e^{1} \wedge e^{2} \wedge e^{
            3} \wedge e^{4}  - e^{1} \wedge e^{2} \wedge e^{5} \wedge e^{6}  +
        e^{3} \wedge e^{4} \wedge e^{5} \wedge e^{6} ).
\end{eqnarray}
$\bullet$ Non-vanishing intrinsic torsion classes:
\begin{eqnarray}
{\cal W}_1 &=& -\frac{2i}{3} \frac{a+b+c}{\sqrt{abc}}, \\
{\cal W}_2 &=& -\frac{4i}{3} \frac{1}{\sqrt{abc}}\left[ a(2a -b-c)
e^{12} - b(2b - a-c)e^{34} + c(2c-a-b)e^{56} \right].
\end{eqnarray}

\subsection*{$SU(2) \times SU(2)$}

$\bullet$ Metric:
\begin{equation}
ds^2=a (\frac{1}{3} e^1 \otimes e^1 + 3 e^6 \otimes e^6) + b
(\frac{1}{3} e^2 \otimes e^2 +
3 e^5 \otimes e^5) + c (\frac{1}{3} e^3 \otimes e^3 +3 e^4 \otimes e^4).
\end{equation}
$\bullet$ $SU(3)$-structure:
\begin{eqnarray}
J &=& a e^{1} \wedge e^{6} - b e^{2} \wedge e^{5} + c e^{3} \wedge
e^{4}, \\
dJ &=& -\frac{1}{\sqrt{3}}(a + b - c) e^{1} \wedge e^{2} \wedge
e^{4} - \frac{1}{\sqrt{3}}(a - b + c)e^{1} \wedge e^{3} \wedge e^{5}\nonumber
\\ &+& \frac{1}{\sqrt{3}}(a - b - c) e^{2} \wedge e^{3} \wedge e^{6}
+
\sqrt{3}(a+b+c) e^{4} \wedge e^{5} \wedge e^{6}, \\
\Omega &=& i\sqrt{\frac{ a b c}{27}}(e^{1} - 3i e^{6}) \wedge (e^{2}
+ 3i e^{5})
\wedge (e^{3} - 3i e^{4}),\\
d\Omega &=& \frac{4 i}{3}\sqrt{abc} ( e^{1} \wedge e^{2} \wedge e^{
            5} \wedge e^{6}  - e^{1} \wedge e^{3} \wedge e^{4} \wedge e^{6}  +
        e^{2} \wedge e^{3} \wedge e^{4} \wedge e^{5} ).
\end{eqnarray}
$\bullet$ Non-vanishing intrinsic torsion classes:
\begin{eqnarray}
{\cal W}_{1} &=& - \frac{2i}{9\sqrt{abc}}(a+b+c),\\
{\cal W}_{2}&=& \frac{4i}{27\sqrt{abc}} (-a(2a-b-c)e^{1}
\wedge e^{6} + b(2b-a-c)e^{2} \wedge e^{5} \nonumber\\
&-& c(2c-a-b)e^{3} \wedge e^{4} ),\\
{\cal W}_{3} &=& - \frac{2}{3\sqrt{3}} ( (a+b-2c)e^{1} \wedge e^{2}
\wedge e^{4}+(a-2b+c) e^{1} \wedge e^{3} \wedge e^{5} \nonumber \\
&+& (-2a+b+c) e^{2} \wedge e^{3} \wedge e^{6}).
\end{eqnarray}

\subsection*{ $Sp(4)/(SU(2)\times U(1))_{non-max}$}

$\bullet$ Metric:
\begin{equation}
ds^2=a (e^1 \otimes e^1 + e^2 \otimes e^2) + b (e^3 \otimes e^3 +
e^4 \otimes e^4) + a (e^5 \otimes e^5 + e^6 \otimes e^6).
\end{equation}
$\bullet$ $SU(3)$-structure:
\begin{eqnarray}
J &=& -a e^{1} \wedge  e^{2} + b e^{3} \wedge e^{4} - a e^{5} \wedge
e^{6},\\
dJ&=& -(2 a + b) (e^{1} \wedge e^{3} \wedge e^{5} + e^{1} \wedge
e^{4} \wedge e^{6} - e^{2} \wedge e^{3} \wedge e^{6} + e^{2} \wedge
e^{4}
\wedge e^{5}),\\
\Omega &=& \sqrt{a^{2} b }(e^{1} + i e^{2}) \wedge (e^{3} - i e^{4}
)
\wedge (e^{5} + i e^{6}),\\
d\Omega &=& 4i \sqrt{a^2b}(e^{1} \wedge e^{2} \wedge e^{3} \wedge
e^{4} - e^{1} \wedge e^{2} \wedge e^{5} \wedge e^{6} + e^{3} \wedge
e^{4} \wedge e^{5} \wedge e^{6}).
\end{eqnarray}
$\bullet$ Non-vanishing intrinsic torsion classes:
\begin{eqnarray}
{\cal W}_1 &=& -\frac{2i}{3} \frac{2a+b}{\sqrt{a^2 b}}, \\
{\cal W}_2 &=& -\frac{4i}{3} \frac{1}{\sqrt{a^2 b}}\left[ a(a -b)
e^{12} - 2b(b - a)e^{34} + a(a-b)e^{56} \right].
\end{eqnarray}

\subsection*{$G_2/SU(3)$}

$\bullet$ Metric:
\begin{equation}
ds^2=a (e^1 \otimes e^1 + e^2 \otimes e^2 + e^3 \otimes e^3 + e^4
\otimes e^4 + e^5 \otimes e^5 + e^6 \otimes e^6).
\end{equation}
$\bullet$ $SU(3)$-structure:
\begin{eqnarray}
J &=& a e^{1} \wedge e^{2} - a e^{3} \wedge e^{4} - a e^{5} \wedge
e^{6},\\
dJ &=& -2 \sqrt{3}a( e^{1} \wedge e^{3} \wedge e^{5}  -
        e^{1} \wedge e^{4} \wedge e^{6} + e^{2} \wedge e^{3} \wedge e^{6} +
        e^{2} \wedge e^{4} \wedge e^{5}),\\
\Omega &=& \sqrt{a^{3}}(e^{1} - i e^{2}) \wedge (e^{3} + i e^{4})
\wedge (e^{5} + i e^{6}),\\
d\Omega &=&  \frac{1}{\sqrt{3}}8i\sqrt{a^{3}} (e^{1} \wedge e^{2}
\wedge e^{3} \wedge e^{4} + e^{1} \wedge e^{2} \wedge e^{5} \wedge
e^{6} - e^{3} \wedge e^{4} \wedge e^{5} \wedge e^{6}).
\end{eqnarray}
$\bullet$ Non-vanishing intrinsic torsion class:
\begin{equation}
{\cal W}_1 = -\frac{4i}{\sqrt{3a}}.
\end{equation}


\providecommand{\href}[2]{#2}\begingroup
\endgroup

\end{document}